\documentclass{elsart}

\usepackage{graphicx}

\usepackage{amssymb}

\begin{document}

\begin{frontmatter}



\title{Perspects in astrophysical databases}


\author[addr1]{Marco Frailis}
\author[addr2]{Alessandro De Angelis}
\author[addr3]{Vito Roberto}

\address[addr1]{Dipartimento di Fisica, Universit\`a di Udine, via delle Scienze 208, 33100 Udine, Italy}
\address[addr2]{INFN, Sezione di Trieste, Gruppo Collegato di Udine, via delle Scienze 208, 33100 Udine, Italy}
\address[addr3]{Dipartimento di Matematica e Informatica, Universit\`a di Udine, via delle Scienze 208, 33100 Udine, Italy}

\begin{abstract}
  Astrophysics has become a domain extremely rich of scientific data. Data
  mining tools are needed for information extraction from such large datasets.
  This asks for an approach to data management emphasizing the efficiency and
  simplicity of data access; efficiency is obtained using multidimensional
  access methods and simplicity is achieved by properly handling metadata.
  Moreover, clustering and classification techniques on large datasets pose
  additional requirements in terms of computation and memory scalability and
  interpretability of results. In this study we review some possible solutions.
\end{abstract}

\end{frontmatter}

\section{Introduction}
\label{introduction}

At present, astrophysics is a discipline in which the exponential growth and
heterogeneity of data require the use of data mining techniques. The primary
source of astronomical data are the systematic sky surveys over a wide energy
range (from $10^{-7}$ eV to $10^{13}$ eV). Large archives and digital sky
surveys with dimensions of $10^{12}$ bytes currently exist, while in the near
future they will reach sizes of the order of $10^{15}$ bytes.  Numerical
simulations are also producing comparable volumes of information.

Several scientific research fields require to perform the analysis on multiple
energy spectra and consequently to get the data from different missions.
Therefore, the use of data mining techniques is necessary to maximize the
information extraction from such a growing quantity of data. This task is
hardened by different issues, like the heterogeneity of astronomical data, due
in part to their high dimensionality including both spatial and temporal
components, due in part to the multiplicity of instruments and projects, or the
use of traditional operational systems, in which the emphasis is on data
normalization, to organize astrophysical data. Data mining for multi-wavelength
analysis necessitates using an informational system, or data warehouse, as a
model for data management, a definition of a common set of metadata to guarantee
the interoperability between different archives and a more efficient data
exploration.

\section{Towards a data whareouse}
Most of the online resources available to the astrophysicists community are
simple data archives containing observational parameters (detector, type of the
observation, coordinates, astronomical object, exposure time, etc.).  Many
astronomical catalogs can be accessed online, but it is still difficult to
correlate objects in different archives or access multiple catalogs
simultaneously. Some advances, in this direction, have been accomplished by
projects like Vizier, Aladin and SkyView \cite{Viz,Sky}.

With an ideal astrophysical database, the users should be able to perform
queries based on scientific parameters (magnitude, redshift, spectral indexes,
morphological type of galaxies, etc.), easily discover the object types
contained into the archive and the available properties for each type, and define
the set of objects which they are interested in by constraining the values of
their scientific properties along with the desired level of detail~\cite{DSZDG00}.

The aforesaid requirements can be satisfied organizing data in a data
warehouse. A data warehouse can be defined as a \emph{subject-oriented},
\emph{integrated}, \emph{time varying} and \emph{non-volatile} data collection
\cite{Inm97}. In a data warehouse, data are arranged in a structure that can be easily
explored and queried, with fewer tables and keys than the equivalent relational
model. You start from a relational model, but some restrictions are introduced by
using \emph{facts}, \emph{dimensions}, \emph{hierarchies} and \emph{measures} in
a characteristic star structure called \emph{star schema}~\cite{Pet94}. The
central table is called ``fact'' table and it is the highest dimensional table of
the scheme. It can represent a particular phenomenon that we want to study. This
table is surrounded by a number of tables, called ``dimensions'', which
represent entities related to the phenomenon to be studied and connected to the
central table, forming the ends of the star. Within the dimensions, attributes
are arranged in hierarchies, determining the ``drill-down'' and ``roll-up''
operations available on each dimension: the result is a tree that the user can
visit from the root to the leaves, refining his query (drill-down) or
generalizing it (roll-up).

Metadata play an important role: a researcher has to obtain information about
the environment in which data have been gathered, in order to understand the
respondence to the project requirements, like date and/or data acquisition
method, internal or external error estimates, aim of data. Computing systems
have to access metadata to merge or compare data from different sources.  For
instance, it is necessary that units are expressed unambiguously to allow
comparisons between data with different units.

The astrophysicists community, in addition to using the FITS (Flexible Image
Transport System) exchange format, is currently considering alternatives like
XML to improve the interoperability. Some attempts to define a common standard
are XSIL (eXtensible Scientific Interchange Language), XDF (eXtensible Data
Format) and VOTable \cite{Sta02}.

\section{Multidimensional access methods}
In the Astroparticle and Astrophysical fields, data is mostly characterized by
multidimensional arrays. For instance, in X-ray and Gamma-ray astronomy, the
data gathered by detectors are lists of detected photons whose properties
include position (RA, DEC), arrival time, energy, error measures both for the
position and the energy estimates (dependent on the instrument response),
quality measures of the events . Source catalogs, produced by the analysis of
the raw data, are lists of point and extended sources characterized by
coordinates, magnitude, spectral indexes, flux, etc.

This multidimensional (spatial) data tend to be large (sky maps can reach sizes
of Terabytes) requiring the integration of the secondary storage, and there is no
total ordering on spatial objects preserving spatial proximity~\cite{GG98}. This
characteristic makes difficult to use traditional indexing methods, like B-trees
or linear hashing.

\emph{Data mining} applied to multidimenisonal data analyzes the relationships
between the attributes of a multidimensional object stored into the database and
the attributes of the neighboring ones. Typical queries required by this kind of
analysis are: \emph{point queries}, to find all objects overlapping the query
point; \emph{range queries}, to find all objects having at least one common
point with a query window; \emph{nearest neighbor queries}, to find all objects
that have a minimum distance from the query object. Another important operation
is the \emph{spatial join}, needed to search multiple source catalogs and
cross-identify sources from different wavebands. Some of the following indexing
methods can be used to improve the queries efficiency.

{\bf HTM}. Data gathered by all sky survays are distributed on an imaginary
sphere. The HTM~\cite{KST01} indexing method maps triangular regions of the
sphere to unique identifiers keeping a certain degree of locality.  The
technique for subdividing the sphere in spherical triangles is a recursive
process.  The starting point is a spherical octahedron which identifies 8
spherical triangles of equal size. In a recursion step, a triangle is further
subdivided into 4 triangles by connecting the side midpoints. At each level of
the recursion, the area of the resulting triangles is roughly the same and each
triangle is uniquely identified by a 2 bit value. This method as been used to
index the Sloan Digital Sky Survay, a catalog of 200 M objects in a
multi-terabyte archive. A level-5 HTM index is used to partition the bulk data.
A database for each level-5 leaf node of the HTM (defining the database file
name) has been built. Each database, containing tuples in a 5-dimensional color
space, is indexed by a KD-tree.

{\bf KD-tree and its variants}. The KD-tree~\cite{Bent75} is a binary tree that
stores points of a $k$-dimensional space. In each internal node, the KD-tree
divides the $k$-dimensional space into two parts with a $(k-1)$-dimensional
hyperplane. The direction of the hyperplane, that is the dimension on which the
division is performed, alternates between the $k$ possibilities from one tree
level to the following one. The subdivision process is recursive and terminates
when the size of a node (its longer side) or the number of points contained into
it is below a certain threshold. Given $N$ data points, the average cost of an
insertion operation is $O(\log_2 N)$. The tree structure and the resulting
hierarchical division of the space depends on the \emph{splitting rule}. A
drawback of KD-trees is that they have to be completely contained into the main
memory. With large datasets this is not feasible.  KD-B-trees~\cite{Rob81} and
hB-trees~\cite{LS90} combine properties of KD-trees and B-trees to overcome this
problem.

{\bf R-tree and its variants}. The R-trees~\cite{Gutt84} are hierarchical
dynamic data structures meant to efficiently index multidimensional objects with
a spatial extent. They are used to store not the real objects but their minimum
bounding box (MBB). Each node of the R-tree corresponds to a disk page. Similar
to B-trees, the R-trees are balanced and they guarantee an efficient memory
usage. Due to the overlapping between the MBBs of sibling nodes, in an R-tree a
range query can require more than one search path to be traversed. Search
performances depend on the insertion algorithms. Some variants have been
proposed to improve the disjointness among regions: the R$^*$-tree \cite{BKSS90}, which
uses a new insertion policy, the SR-tree \cite{KS97}, which uses the intersection of
bounding spheres and bounding rectangles to keep small the diameters and volumes
of the regions, and the A-tree \cite{SYUK02}, which improves the fanout of the nodes
using an approximation of the MBRs.

Usually, the analysis of astrophysical data is performed on a static dataset.
In this case, an optimized index (in terms of memory and query performances) can
be built using a priori information on the dataset. Several bulk loading
techniques have been proposed in the literature. We have followed a top-down
construction method called VAMSplit algorithm, described in \cite{WJ96}, to
build and optimized R-Tree. The main idea is to find a split strategy that
minimizes the number of buckets used and provides a good query performance. This
is achieved by recursively splitting the dataset on a near median element along
the dimension with maximum variance. To adapt it to a large dataset, we had to
implement an external selection algorithm. The implementation uses a sampling
method suggested by \cite{MR01} to find a good pivot value and reduce the number
of I/O operations; a caching strategy explained in \cite{BK99} has been adopted
to partition the data into the secondary memory.

\section{Clustering algorithms on large datasets}
Clustering algorithms have to locate regions of interest in which to perform more
detailed analysis and point out correlations between objects. An important
issue, in large datasets, is the efficiency and scalability of the clustering
algorithms with respect to the dataset size.

Many scalable algorithms have been proposed in the last ten years, including:
BIRCH, CURE, CLIQUE~\cite{Ber02}.

In particular, BIRCH is a hierarchical clustering algorithm. The main idea
behind the algorithm is to compress data into small subclusters and then to
perform a standard partitional clustering on the subclusters. Each subcluster is
represented by a \emph{clustering feature} which is a triplet summarizing
information about the group of data objects, that is the number of points
contained into the cluster and the linear sum and the square sum of the data
points. This algorithm has a linear cost with respect to the number of data
points.

CURE is an hierarchical agglomerative algorithm. Instead of using a single
centroid or object, it selects a fixed number of well-scattered objects to
represent each cluster. The distance between two clusters is defined as the
distance between the closest pair of representatives points and at each step of the
algorithm, the two closest clusters are merged. The algorithm terminates when
the desired number of clusters is obtained. To reduce the computational cost of
the algorithm, these steps are performed on a data sample (using suitable
sampling techniques). Its computational cost is not worse than the BIRCH one.

CLIQUE has been designed to locate clusters in subspaces of high dimensional
data. This is useful because generally, in high dimensional spaces, data are
scattered. CLIQUE partitions the space into a grid of disjoint rectangular units
of equal size. The algorithm is made up of three phases: first, it finds
subspaces containing clusters of dense units, than identifies the clusters,
and finally generates a minimum description for each cluster. Also this
algorithm scales linearly with the database size.

\section{Novelty detection: Support Vector Clustering}
Support Vector Machines and the related kernel methods are becoming popular for
data mining tasks. In many real problems, the task is not classifying but
novelties or anomalies detecting. In astrophysics, possible applications are the
research of anomalous events or new astronomical sources. An approach is finding
the \emph{support} of a distribution (rather than estimating the density
function of the data), thus avoiding the need of an a priori parameterized model
of the distribution. A method to solve this problem is represented by the
Support Vector Clustering (SVC) algorithm~\cite{BHSV01}, in which data are
mapped to a higher dimensional space by means of a Gaussian kernel function. In
the new space, the algorithm finds the minimum sphere enclosing the data. The
mapping of the sphere to the original input space generates a set of contours
enclosing the data and corresponding to the support of the
distribution. Outliers are defined as the Bounding Support Vectors (BSV).

\section{Conclusions}
In this work we have studied some data management and mining issues related to
astrophysical data, aiming at a complete data mining framework. In particular,
we have justified the need for a data warehousing approach to handle
astrophysical data and we have focused on multidimensional access methods to
efficiently index spatial and multidimensional data. A second issue concerns
clustering techniques on large datasets, and we have discussed about some
scalable algorithms with linear computational complexity. Finally, we have
outlined the usefulness of non-parametric clustering algorithms, like the SVC,
for novelty detection.



\end{document}